\documentclass{eas}
\usepackage{graphicx}
\usepackage{color}
\usepackage{amsmath,amssymb}

\newcommand{\aap}{A\&A}  %{\textit{Astronomy \& Astrophysics}}
\newcommand{\aj}{AJ}      %{\textit{Astronomical Journal}}
\newcommand{\apjl}{ApJL}   %{\textit{Astrophysical Journal Letters}}
\newcommand{\mnras}{MNRAS} %{\textit{Monthly Notices of the R.A.S.}}
\newcommand{\pasj}{PASJ}   %{\textit{Publications of the Astronomical Society of Japan}}
\newcommand{\msun}{\ensuremath{\mathrm{M}_{\odot}}}
\newcommand{\kms}{\ensuremath{\mathrm{km}\,\mathrm{s}^{-1}}}

\begin{document}

%%-----------------------------
%%      the top matter
%%-----------------------------
\title{Numerical models for the circumstellar medium around Betelgeuse} 
\runningtitle{Mackey \etal{}: Betelgeuse's evolving circumstellar medium}
\author{Jonathan Mackey}\address{Argelander-Institut f\"ur Astronomie, Auf dem H\"ugel 71, 53121 Bonn, Germany \email{jmackey@astro.uni-bonn.de}}
\author{Shazrene Mohamed}\address{South African Astronomical Observatory, P.O.\ Box 9, Observatory, Cape Town, 7935, South Africa}
\author{Hilding R.\ Neilson}\address{
  Department of Physics \& Astronomy,
  East Tennessee State University,
  Box 70652, Johnson City, TN, 37614, USA}
\author{Norbert Langer}\sameaddress{1}
\author{Dominique M.-A.\ Meyer}\sameaddress{1}
\begin{abstract}
The nearby red supergiant (RSG) Betelgeuse has a complex circumstellar medium out to at least 0.5 parsecs from its surface, shaped by its mass-loss history within the past $\approx0.1$ Myr, its environment, and its motion through the interstellar medium (ISM).
In principle its mass-loss history can be constrained by comparing hydrodynamic models with observations.
Observations and numerical simulations indicate that Betelgeuse has a very young bow shock, hence the star may have only recently become a RSG.
To test this possibility we calculated a stellar evolution model for a single star with properties consistent with Betelgeuse.
We incorporated the resulting evolving stellar wind into 2D hydrodynamic simulations to model a runaway blue supergiant (BSG) undergoing the transition to a RSG near the end of its life.
The collapsing BSG wind bubble induces a bow shock-shaped inner shell which at least superficially resembles Betelgeuse's bow shock, and has a similar mass.
Surrounding this is the larger-scale retreating bow shock generated by the now defunct BSG wind's interaction with the ISM.
We investigate whether this outer shell could explain the bar feature located (at least in projection) just in front of Betelgeuse's bow shock.
\end{abstract}
\maketitle

%%-----------------------------
%%      your text
%%-----------------------------

% -------------------------------------------------------------------
\section{Introduction} \label{sec:intro}
% -------------------------------------------------------------------
Betelgeuse ($\alpha$ Orionis) is one of the brightest stars on the sky and one of the nearest red supergiants (RSG), but its circumstellar medium (CSM) has only begun to reveal its secrets in the past 20 years.
Part of the reason for this is that optical and near-infrared observations are very difficult because the star is very bright at these wavelengths, and so most detections of the CSM are from space-based far-infrared and ultra-violet observatories.
\textit{IRAS} observations (Noriega-Crespo \etal{} \cite{NorBurCaoEA97}) of Betelgeuse detected a bow shock and a linear ``bar'' feature upstream from the bow shock and perpendicular to the star's proper motion.
Later \textit{AKARI} observations (Ueta \etal{} \cite{UetIzuYamEA08}) confirmed these features with higher resolution observations of part of the bow shock.
\textit{Herschel} observations (Cox \etal{} \cite{CoxKervMarEA12}; Decin \etal{} \cite{DecCoxRoyEA12}) have dramatically better spatial resolution than previous data and resolve the bow shock into a number of layers.
\textit{Herschel's} better wavelength coverage also allows more detailed dust modelling, providing mass estimates for Betelgeuse's circumstellar structures (Decin \etal{} \cite{DecCoxRoyEA12}).
A thin shell of far-ultraviolet emission was found by Le Bertre \etal{}~(\cite{LeBMatGerEA12}) at the same position as the IR bow shock in archival \textit{GALEX} data.
They also presented VLA 21cm observations with evidence of some atomic hydrogen near the bow shock, and stronger emission from an inner shell three times closer to Betelgeuse.

Mohamed, Mackey, \& Langer~(\cite{MohMacLan12}) presented the first 3D hydrodynamical simulations of Betelgeuse's bow shock, assuming a constant wind.
They found that for a wide range of parameters the bow shock should have a mass $M\approx 0.05-0.15 \,\msun$, more than $10\times$ larger than the mass $M\approx0.0033\,\msun$ they estimated from \textit{AKARI} data.
This result was confirmed in the 2D bow shock simulations of Decin \etal{}~(\cite{DecCoxRoyEA12}), and the discrepancy is even more severe when one considers the best mass estimate from \textit{Herschel} of $M\approx0.0024\,\msun$ (Decin \etal{}~\cite{DecCoxRoyEA12}).

Mohamed \etal{}~(\cite{MohMacLan12}) concluded that the bow shock must be very young ($\lesssim30$ kyr) and not yet in a steady state.
This means that either the stellar wind or the ISM properties have changed dramatically on this timescale.
They further suggested that if Betelgeuse had recently evolved from a blue supergiant (BSG) to a RSG, then the upstream bar could be a remnant bow shock from an earlier main sequence or BSG wind, currently being overtaken by the star.
In Mackey \etal{}~(\cite{MacMohNeiEA12}) we tested this idea by simulating the CSM produced by a runaway 15\,\msun{} star as it evolves from a BSG to a RSG at the end of core helium-burning.
The results of Mackey \etal{}~(\cite{MacMohNeiEA12}) are summarised in this paper, and are further discussed in the context of the new results from Decin \etal{}~(\cite{DecCoxRoyEA12}).
Section~\ref{sec:methods} describes the stellar evolution model and simulation setup, section~\ref{sec:results} summarises our results, and section~\ref{sec:discussion} presents some discussion and conclusions.

% -------------------------------------------------------------------
\section{Description of Model} \label{sec:methods}
\begin{figure}
\centering
\resizebox{0.7\hsize}{!}{\includegraphics{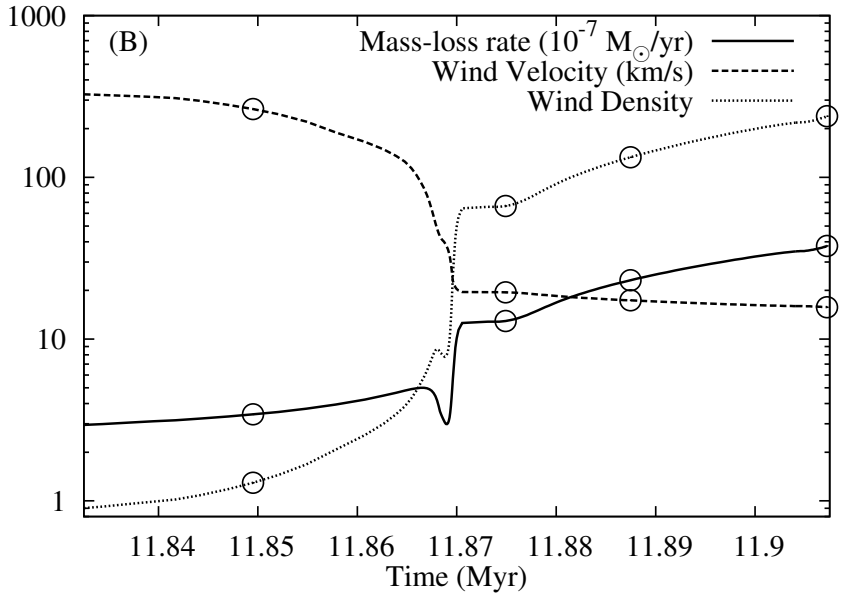}}
\caption{The wind mass-loss rate, $\dot{M}$ in $10^{-7}\ \msun\,\mathrm{yr}^{-1}$, wind velocity, $v_w$ in $\mathrm{km}\,\mathrm{s}^{-1}$, and  $\dot{M}/v_w$ (proportional to wind density, arbitrary units) are shown.
  Open circles identify the times of the four snapshots in Fig.~\ref{fig:DensTempVel}.
  }
\label{fig:WIND}
\end{figure}

% -------------------------------------------------------------------
A $15\ \msun$ solar metallicity star is modelled travelling at $v_\star=50\ \kms$ through the diffuse ISM with hydrogen number density $n_{\mathrm{H}}=0.2\ \mathrm{cm}^{-3}$ and temperature $T=7700$ K.
Stellar evolution is calculated using the Yoon \& Langer~(\cite{YooLan05}) stellar evolution code including mass loss but no convective core overshooting.
The tip of the RSG branch in this model has properties consistent with Betelgeuse (Neilson \etal{}~\cite{NeiLesHau11}, and references therein).
Wind velocities are approximated using the Eldridge \etal{}~(\cite{EldGenDaiEA06}) prescription, $v_w^2 =  \beta_w(T) v_{\mathrm{esc}}^2$.
To approximate the slow wind of Betelgeuse we use $\beta_w=0.04$ for $T<3600$ K (instead of the recommended value of 0.125).
After the main sequence the star becomes a RSG; it then embarks on an extended blue loop for most of the He-burning phase; and finally evolves again to a RSG for the last $\approx40$ kyr of its life.
The wind velocity,  mass-loss rate $\dot{M}$, and wind density ($\dot{M}/v_w$), are plotted in Fig.~\ref{fig:WIND} for the last $\approx75$ kyr of evolution, corresponding to the end of the blue loop and evolution along the RSG branch.

The circumstellar medium is calculated by solving the inviscid Euler equations for an ideal gas on a uniform axisymmetric grid in $(z,R)$ in the rest frame of the star.
A second-order-accurate integration scheme is used, and radiative cooling is included as a source term to the energy equation using the collisional ionisation equilibrium cooling curve of Wiersma \etal{}~(\cite{WieSchSmi09}).
The code is described in more detail in Mackey \etal{}~(\cite{MacMohNeiEA12}) and Mackey \& Lim~(\cite{MacLim10}).
The stellar wind boundary condition is implemented by imposing a freely expanding wind within a 20 grid-zone radius of the origin, updating the wind parameters from the stellar evolution model every timestep.

% -------------------------------------------------------------------
\section{Results} \label{sec:results}
\begin{figure}
\centering
\resizebox{1.0\hsize}{!}{\includegraphics{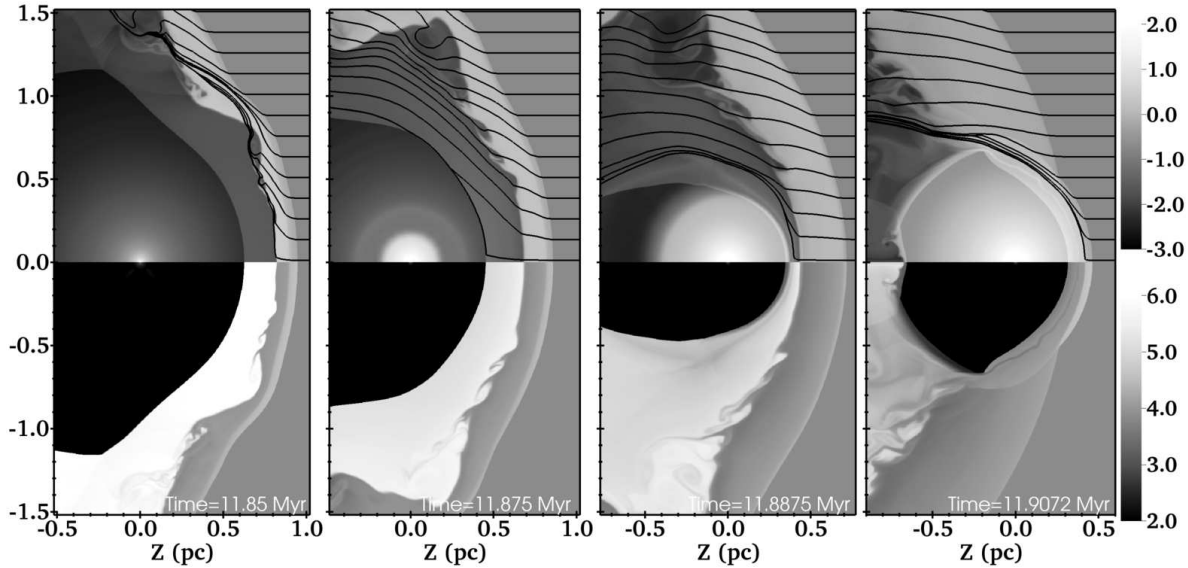}}
\caption{
  Log of hydrogen number density $n_{\mathrm{H}}$ (upper half-plane, in $\mathrm{cm}^{-3}$) and log of temperature (lower half-plane, in Kelvin) plotted on logarithmic scales at times $t=11.85,\ 11.875,\ 11.8875,$ and $11.9073$ Myr from left to right, respectively.
  Velocity streamlines are also plotted in the upper half-plane.
  Only part of the simulation domain is shown.
  The snapshot times correspond to the circled times in Fig.~\ref{fig:WIND}.
  }
\label{fig:DensTempVel}
\end{figure}

% -------------------------------------------------------------------
Snapshots from the simulation are shown in Fig.~\ref{fig:DensTempVel} at the four evolutionary times marked by circles in Fig.~\ref{fig:WIND}.
Log of $n_{\mathrm{H}}$ is shown in the upper half-plane together with velocity streamlines traced from the right-hand boundary.
The lower half-plane shows log of $T$, where the rapidly decelerating wind is seen by the decreasing temperature in the shocked-wind region from $T>10^6$ K initially to $T<10^4$ K in the last snapshot.

The first snaphot shows the CSM at the end of the BSG evolutionary phase, with a single bow shock comprising (from small radii to large):
the freely expanding wind out to $r\approx0.5$ pc;
the reverse shock;
the hot shocked wind bubble at lower density than the ambient ISM;
the contact discontinuity distorted by Kelvin-Helmholz instability;
the cooled and shocked ISM shell; and finally
the forward shock at about $r=0.95$ pc along the symmetry axis.

In the second snapshot the star has become a RSG and hence the hot bubble has lost its energy source, so it is cooling and expanding back towards the star.
The streamlines show that the contact discontinuity no longer has strong shear across it because it has also begun receding to the star.

By $t=11.8875$ Myr in the third snapshot, the RSG wind has begun interacting with the collapsing hot bubble, generating an inner shell around the upstream hemisphere of the RSG wind that is shaped like a bow shock.
The final snapshot shows the CSM at the pre-supernova stage.
Here the hot gas from the BSG wind bubble is all downstream from the star, the RSG wind is shock-confined (but still advancing) on all sides, and the upstream shock is now advancing into undisturbed ISM.

\begin{figure}[t]
\centering
\resizebox{0.8\hsize}{!}{\includegraphics{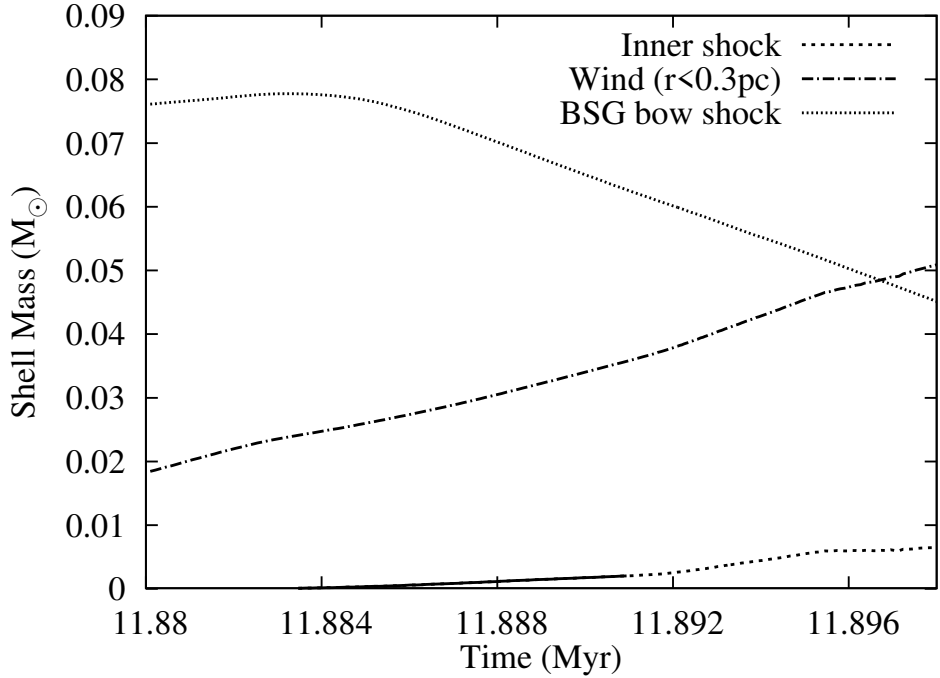}}
\caption{Masses of CSM structures as the simulation evolves.
  The green solid curve shows the total mass in the BSG bow shock upstream from the star.
  The blue solid line shows the total mass of freely-expanding wind within 0.3 pc of the star.
  The black solid and dotted line shows the mass of the inner shocked-shell as a function of time in the region upstream from the star.
  At $t\approx11.891$ Myr the inner shocked-shell begins to merge with the contact discontinuity from the outer shell; this merged overdensity is shown with the dotted line.
  }
\label{fig:ShellMass}
\end{figure}

In Mackey \etal{}~(\cite{MacMohNeiEA12}) we suggested that the inner shocked-shell in the third panel of Fig.~\ref{fig:DensTempVel} could correspond to the bow shock seen at $r\approx0.35$ pc around Betelgeuse, and the receding BSG bow shock is a plausible candidate structure for the upstream bar at $r\approx0.45$ pc.
The masses of the unshocked RSG wind, of the shocked ISM from the relic BSG bow shock, and of the inner shocked-shell are plotted as a function of time in Fig.~\ref{fig:ShellMass} during the time when the inner shocked-shell exists.
The RSG wind mass increases almost linearly to $M\approx0.05\ \msun$ over this timescale, and at the same time the mass of shocked ISM from the BSG bow shock decreases to about the same value as it is swept downstream (only shocked ISM upstream from the star is counted).
The inner shocked-shell is much smaller than both of these, growing from zero to $M\approx0.002\ \msun$ at 11.891 Myr, when it reaches the upstream contact discontinuity and begins to merge with it.

% -------------------------------------------------------------------
\section{Discussion} \label{sec:discussion}
% -------------------------------------------------------------------
\subsection{Comparison with recent observations}
Decin \etal{}~(\cite{DecCoxRoyEA12}) have carefully analysed the \textit{Herschel} Betelgeuse observations
with detailed dust modelling, to estimate the masses contained in the bow shock and bar.
We compare here with their best estimates for the masses, but note that the uncertainties in the estimates are at the order-of-magnitude level.
They obtain a bow shock total mass of $M\approx0.0024\ \msun$, in good agreement with the estimate of $M\approx0.0033\ \msun$ by Mohamed \etal{}~(\cite{MohMacLan12}) from \textit{AKARI} data.
This is marginally larger than the mass of the inner shocked-shell in our simulation, but given the uncertainties this is very good agreement.

If the linear bar is composed of ISM gas (which it is in our model), with ISM dust content, Decin \etal{}~(\cite{DecCoxRoyEA12}) calculate its mass to be $M\approx0.01\ \msun$.
This is $5-7\times$ smaller than the remnant BSG bow shock in our simulations (see Fig.~\ref{fig:ShellMass}), but in our model only the part of this mass closest to the star would be emitting brightly in the far-infrared.
The main argument against our model from the \textit{Herschel} data is that the bar seems to have no curvature, whereas our simulation shows some curvature.
The rotational symmetry of the simulations presented here significantly limits the bow shock shape however; in particular it can only be perfectly linear if it is aligned exactly perpendicular to the motion.
Three-dimensional simulations that allow the bow shock to move in all directions are needed to determine whether such a linear feature would naturally arise from a circumstellar bow shock.

The mass of the RSG wind was measured with 21cm neutral H observations by Le Bertre \etal{}~(\cite{LeBMatGerEA12}), and they obtained $M\approx0.068\ \msun$.
With a different neutral H dataset, Decin \etal{}~(\cite{DecCoxRoyEA12}) derive a similar mass of $M\approx0.02-0.07\ \msun$.
These measurements are within a factor of 2 of the RSG wind mass in our simulations, although it is something of a mystery that Le Bertre \etal{}~(\cite{LeBMatGerEA12}) see evidence that the neutral H is confined to a radius $r<0.12$ pc.
Nothing in our simulations provides such a confinement.

The \textit{GALEX} far-ultraviolet emission (Le Bertre \etal{}~\cite{LeBMatGerEA12}) at the same position as the IR bow shock may further support our model.
If this emission is excited by hot electrons, the remnant BSG hot bubble is a natural place to find them, and they would interact directly with the outer edge of the expanding RSG wind.
Further work is required in order to make a quantitative prediction of the far-ultraviolet emission from our simulations.

\subsection{Summary}
We have presented a new model for the CSM around Betelgeuse, in which the star has recently evolved from a BSG to a RSG and is now within 20 kyr of the end of its life.
The evolving stellar wind together with the stellar motion produce a complex CSM with multiple shells and even regions of hot gas with $T>10^5$ K within 1 pc of the star (especially downstream).
The inner shocked-shell around the expanding RSG wind has a comparable mass and radius to the observed bow shock around Betelgeuse, and is $20-50\times$ less massive than an equilibrium RSG bow shock of a similar size.

The remnant of the BSG bow shock forms a large structure upstream from the inner shocked-shell, with a much larger radius of curvature because it originally had a much larger standoff distance from the star.
Its location makes it a plausible candidate for the bar upstream from Betelgeuse's bow shock, and its mass seems to have the right order-of-magnitude (although this is very uncertain).
Its curvature does not agree with the observed bar, which has no discernable curvature, but this is not robustly predicted by our two-dimensional simulations.

The model we present therefore has some encouraging agreement with observations, in particular in explaining the very low mass of the bow shock, which is difficult to explain in any other way.
On the other hand some questions remain, notably the apparent confinement of the RSG wind to $r<0.12$ pc from neutral H observations, and also the detailed structure of the linear bar.
Further observational study of the CSM is therefore required, in particular data that measure the gas as well as the dust, because the dust-to-gas mass conversion is very uncertain and prevents us placing strong constraints on our model.
Spectral data that constrain the gas velocity and its chemical state would also be very useful, although the star is so bright that this is difficult at optical wavelengths.

\section*{Acknowledgements}
JM and HN acknowledge funding from the Alexander von Humboldt Foundation.
Simulations were performed on the JUROPA supercomputer at J\"ulich Supercomputing Centre (project HBN23).
This work was supported by the Deutsche Forschungsgemeinschaft priority program 1573, ``Physics of the Interstellar Medium''.

\end{document}